\documentclass{ckm}                 

\usepackage{txfonts}            
\confname{Workshop on the CKM Unitarity Triangle, IPPP Durham, April
  2003}

\newcommand{\be}{\begin{equation}}
\newcommand{\ee}{\end{equation}}
\newcommand{\bea}{\begin{eqnarray}}
\newcommand{\eea}{\end{eqnarray}}
\newcommand{\no}{\nonumber}

\newcommand{\cO}{{\cal O}}

\def\arnps#1#2#3{  {\it Annu. Rev. Nucl. Part. Sci. }{\bf #1}, #3 (#2)}
\def\npb#1#2#3{    {\it Nucl. Phys.}~B {\bf #1}, #3 (#2)}
\def\plb#1#2#3{    {\it Phys. Lett.}~B {\bf #1}, #3 (#2)}

\def\prd#1#2#3{    {\it Phys. Rev.}~D {\bf #1}, #3 (#2)}

\def\prl#1#2#3{    {\it Phys. Rev. Lett. }{\bf #1}, #3 (#2)}

\def\ijmpa#1#2#3{  {\it Int. J. Mod. Phys. }{\bf A #1}, #3 (#2)}

\def\jhep#1#2#3{   {\it JHEP  }{\bf #1}, #3 (#2)}

\def\ibid#1#2#3{{\bf #1}, #3 (#2)}

\title{$K\to\pi\nu\bar\nu$ decays and CKM fits}

\author{Gino Isidori}
\address{INFN, Laboratori Nazionali di Frascati, I-00044 Frascati, Italy}

\begin{document}

\begin{abstract}
After a brief introduction to the so-called flavour problem,
we discuss the role of $K \to \pi \nu \bar{\nu}$ decays 
in shedding new light on this issue. In particular,
we review the theoretical uncertainties in predicting 
$\Gamma(K \to \pi \nu {\bar\nu})$ within the SM, 
the sensitivity of these observables to 
New Physics scenarios, and the status of their 
experimental determination.
\end{abstract}

\maketitle

\vskip 0.7 cm 

\section{Introduction: the flavour problem}
There is no doubt that the Standard Model (SM) provides a successful 
and economical description of particle physics up to 
energies of $\cO(100~{\rm GeV})$. However, 
it is very natural to consider this model only as the low-energy limit 
of a more general theory, or as the renormalizable 
part of an effective field theory valid up to some 
still undetermined cut-off scale $\Lambda$. 
We have no direct indications about the value of this cut-off, 
but theoretical arguments based on a natural solution of the 
hierarchy problem suggest that $\Lambda$ should not exceed a few TeV. 

From this perspective, the goal of indirect New Physics (NP) searches
can be viewed as the search for the effective non-renormalizable 
interactions, suppressed by inverse powers of $\Lambda$, 
which encode the presence of new degrees of freedom
at high energies. These operators should naturally induce large effects 
in processes which are not mediated by tree-level SM amplitudes, 
such as $\Delta F=1$ and  $\Delta F=2$
flavour-changing neutral current (FCNC) transitions. 
Up to now there is no evidence of these effects and this 
implies severe bounds on the effective scale of several 
dimension-six operators (more than $100$~TeV for the effective 
scale of the $\Delta S=2$ operators 
contributing to  $K^0$--${\bar K}^0$ mixing).
The apparent contradiction between these high bounds on $\Lambda$
and the expectation  $\Lambda \sim$~TeV, dictated by the electroweak
hierarchy problem, is a manifestation of what in 
many specific NP frameworks
goes under the name of {\em flavour problem}.

In the last few years  the flavour problem has been considerably exacerbated 
by the new precise data in the $B$ system, which show an excellent
consistency of the various observables used to (over-)constrain 
the CKM unitarity triangle \cite{CKMBook}.
One could therefore doubt about the need for new precision 
measurements. 
However, the present consistency of CKM fits 
should not be over emphasized and there are various 
reasons why a deeper study of FCNCs and, particularly, 
of rare $K$ decays, would still be very useful.

First of all, in order 
to constraint the parameter space of possible 
SM extensions, we cannot simply test the consistency of 
the SM hypothesis, as is usually done in present 
CKM fits. In principle, all observables potentially 
sensitive to NP, namely all short-distance dominated 
FCNCs, should be left as free parameters. 
In other words, we should try to perform in the 
flavour sector something similar to what has been done 
in the electroweak sector with the model-independent 
fits of the oblique corrections (see e.g.~Ref.~\cite{Alt}). 
A completely model-independent approach to the flavour 
problem is very difficult, because of the larger number 
of couplings involved. Nonetheless, we can already 
try to constrain the parameter space of a series of 
rather general NP frameworks, such as 
\begin{enumerate}
\item{} models with Minimal Flavour Violation \cite{MFV1,MFV2,MFV};
\item{} models with large NP effects in $b \to s$ FCNC transitions
and not in $b \to d$ and  $b \to s$ ones (or permutations) \cite{Nir};
\item{} models with large NP effects only in $\Delta F=2$ 
        FCNC amplitudes \cite{GNW,Silva,Quim}; 
\item{} models with large NP effects only in $Z$-penguins 
        FCNC amplitudes \cite{Zpenguins};
\end{enumerate} 
and a few other cases of well-defined effective field theories.
As can by easily understood, in all these scenarios a substantial 
progress with respect to the present situation would be obtained 
by the inclusion of the precise $\Delta F=1$ constraint
from $K\to \pi\nu\bar\nu$ decays (see e.g. Refs.~\cite{MFV,rising}).

A second important argument if favour of precise 
measurements of  $K\to \pi\nu\bar\nu$ widths,  
is the fact that most of the observables used in present 
CKM fits, such as $\epsilon_K$, $\Gamma(b\to u \ell \bar{\nu})$ or $\Delta M_{B_d}$,
suffer from irreducible theoretical errors at the 
10$\%$ level (or above). In the perspective of reaching 
a high degree of precision, it would be desirable to 
base these fits only on observables with theoretical errors
at the percent level (or below), such as the CP asymmetry in $B\to J/\Psi K_S$. 
As we shall review in the following section, 
the $K_L \to \pi^0 \nu\bar\nu$ width belongs to this category.

\section{Theoretical predictions of $\Gamma(K \to \pi\nu\bar\nu)$}

\begin{figure}[t]
$$
\includegraphics[width=7.5cm,height=2.8cm]{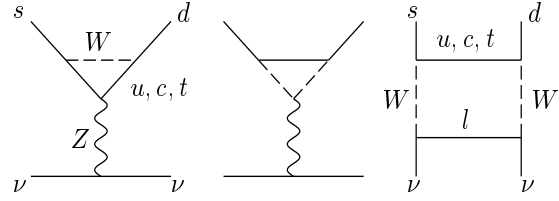}
$$
\caption{One-loop diagrams contributing to the $s \to d \nu \bar{\nu}$ transition.}
\label{fig:Kpnn}
\end{figure}

The $s \to d \nu \bar{\nu}$ transition is one 
of the rare examples of weak processes whose 
leading contribution starts at ${\cal O}(G^2_F)$. At the one-loop 
level it receives contributions only from $Z$-penguin and 
$W$-box diagrams, as shown in Fig.~\ref{fig:Kpnn}, 
or from pure quantum electroweak effects.
Separating the contributions to the one-loop amplitude according to the 
intermediate up-type quark running inside the loop, we can write
\be
{\cal A}(s \to d \nu \bar{\nu}) = \sum_{q=u,c,t} V_{qs}^*V_{qd} {\cal A}_q~,  
\ee
where $V_{ij}$ denote the elements of the CKM matrix. 
The hierarchy of these elements 
would favour  up- and charm-quark contributions;
however, the {\em hard} GIM mechanism of the perturbative calculation
implies ${\cal A}_q \sim m^2_q/M_W^2$, leading to a completely 
different scenario. The top-quark contribution turns out to be 
the leading term both in the real 
and in the imaginary part of the amplitude.
This structure implies several interesting consequences for
${\cal A}(s \to d \nu \bar{\nu})$: it is dominated by short-distance
dynamics, therefore its QCD corrections are small and calculable in perturbation theory; 
it is very sensitive to $V_{td}$, which is one of the less constrained CKM matrix elements;
it is likely to have a large CP-violating phase; it is very suppressed within the SM and thus 
very sensitive to possible new sources of quark-flavour mixing.

Short-distance contributions to ${\cal A}(s \to d \nu \bar{\nu})$,
are efficiently described, within the SM, by the 
following effective Hamiltonian \cite{BB2}
\bea
&& {\cal H}_{eff} = \frac{G_F}{\sqrt 2} \frac{\alpha}{2\pi \sin^2\Theta_W}
 \sum_{l=e,\mu,\tau} \left[ \lambda_c X^l_{NL} + \lambda_t X(x_t) \right] \no \\
&& \qquad \times (\bar sd)_{V-A}(\bar\nu_l\nu_l)_{V-A}~,
\label{eq:Heff} 
\eea
where $x_t=m_t^2/M_W^2$ and, as usual, $\lambda_q = V^*_{qs}V_{qd}$. 
The coefficients $X^l_{NL}$ and 
$X(x_t)$, encoding top- and charm-quark loop contributions, 
are known at the NLO accuracy in QCD \cite{BB,MU} and can be found 
explicitly in \cite{BB2}. The theoretical uncertainty in the dominant 
top contribution is very small and it is essentially determined by the 
experimental error on $m_t$. Fixing the $\overline{\rm MS}$ top-quark mass 
to ${\overline m}_t(m_t) = (166 \pm 5)$~GeV we can write
\be
X(x_t) = 1.51 \left[ \frac{ {\overline m}_t(m_t)}{166~\rm GeV} \right]^{1.15} = 
1.51 \pm 0.05~.
\ee
The simple structure of ${\cal H}_{eff}$ leads to two important  
properties of the physical $K\to \pi \nu\bar \nu$ transitions:
\begin{itemize}
\item{} The relation between partonic and hadronic amplitudes 
is exceptionally accurate, since hadronic matrix elements
of the $\bar{s} \gamma^\mu d$ current between a kaon and a pion
can be derived by isospin symmetry from the measured $K_{l3}$ rates. 
\item{} The lepton pair is produced in a state of definite CP 
and angular momentum, implying that the leading SM contribution 
to $K_L \to \pi^0  \nu \bar{\nu}$ is CP-violating.
\end{itemize}

\medskip\noindent
The largest theoretical uncertainty 
in estimating ${\cal B}(K^+\to\pi^+ \nu\bar{\nu})$ 
originates from the charm sector.  
Following the analysis of Ref.~\cite{BB2},
the perturbative charm contribution is 
conveniently described in terms of the parameter 
\be
 P_0(X) = \frac{1}{\lambda^4}
\left[\frac{2}{3}X^e_{NL}+\frac{1}{3}X^\tau_{NL}\right] = 0.39 \pm 0.06~,
\label{eq:P0}
\ee 
where we have used $\lambda = 0.2240 \pm 0.0036$ \cite{CKMBook}.
The numerical error in the r.h.s. of Eq.~(\ref{eq:P0})  is obtained 
from a conservative estimate of NNLO corrections \cite{BB2}. 
Recently also non-perturbative effects introduced 
by the integration over charmed degrees of freedom
have been discussed \cite{Falk_Kp}. Despite a precise 
estimate of these contributions is not possible at present
(due to unknown hadronic matrix-elements), these can be 
considered as included in the uncertainty quoted in 
Eq.~(\ref{eq:P0}).\footnote{~The natural order of magnitude 
of these non-perturbative corrections, 
relative to the perturbative charm contribution 
is $m_K^2/(m_c^2 \ln(m^2_c/M^2_W)) \sim 2 \%$.}
Finally, we recall that genuine long-distance effects 
associated to light-quark loops are well below  
the uncertainties from the charm sector \cite{LW}.

With these definitions the branching fraction of $K^+\to\pi^+\nu\bar\nu$ 
can be written as   
\bea
&& {\cal B}(K^+\to\pi^+\nu\bar\nu) = \no \\
&&  \frac{ \bar \kappa_+ }{ \lambda^2 } ~
\left[ ({\rm Im} \lambda_t)^2 X^2(x_t) +
\left( \lambda^4 {\rm Re}\lambda_c  P_0(X)+
       {\rm Re}\lambda_t X(x_t)\right)^2 \right]~, \no\\
\label{eq:BRSM} 
\eea
where \cite{BB2}
\be 
\bar \kappa_+ = r_{K^+} \frac{3\alpha^2 {\cal B}(K^+\to\pi^0 e^+\nu)}{
2\pi^2\sin^4\Theta_W} = 7.50 \times 10^{-6} 
\ee
and $r_{K^+}=0.901$ takes into account the isospin breaking corrections 
necessary to extract the matrix element of the $(\bar s d)_{V}$ current 
from ${\cal B}(K^+\to\pi^0 e^+\nu)$ \cite{MP}. 

The case of $K_L\to\pi^0 \nu\bar{\nu}$ is even cleaner from the
theoretical point of view \cite{Litt}.  
Because of the  CP structure, only the imaginary parts in (\ref{eq:Heff}) 
--where the charm contribution is absolutely negligible--
contribute to ${\cal A}(K_2 \to\pi^0 \nu\bar{\nu})$. Thus 
the dominant direct-CP-violating component 
of ${\cal A}(K_L \to\pi^0 \nu\bar{\nu})$ is completely 
saturated by the top contribution, 
where  QCD corrections are suppressed and rapidly convergent. 
Intermediate and long-distance effects in this process
are confined only to the indirect-CP-violating 
contribution \cite{BB3} and to the CP-conserving one \cite{CPC},
which are both extremely small.
Taking into account the isospin-breaking corrections to the 
hadronic matrix element \cite{MP}, we can write an
expression for the $K_L\to\pi^0 \nu\bar{\nu}$ rate in terms of 
short-distance parameters, namely
\bea
&& {\cal B}(K_L\to\pi^0 \nu\bar{\nu})_{\rm SM} =
 \frac{ \bar \kappa_L }{ \lambda^2 } ({\rm Im} \lambda_t)^2 X^2(x_t) \\
&& = 1.48 \times 10^{-11} \times \left[
\frac{\overline{m}_t(m_t) }{ 166~{\rm GeV}} \right]^{2.30} \left[ 
\frac{{\rm Im} \lambda_t }{ 10^{-4} } \right]^2~, 
\eea
which has a theoretical error below $3\%$.

At present the SM predictions of the two  $K\to \pi\nu\bar\nu$ rates
are not extremely precise owing to the limited knowledge of both 
real and imaginary parts of $\lambda_t$. 
Taking into account the latest input values reported 
in Ref.~\cite{CKMBook} and the corresponding global
fit of the CKM unitarity triangle, we find 
${\rm Re} \lambda_t = -(3.11 \pm 0.21)\times 10^{-4}$
and   ${\rm Im} \lambda_t = (1.33 \pm 0.12)\times 10^{-4}$,
which yield to
\bea
{\cal B}(K^+ \to\pi^+ \nu\bar{\nu})^{ }_{\rm SM} &=& (0.77 \pm 0.11) 
\times 10^{-10}~, \qquad  \label{BRK+nnt}
\\
{\cal B}(K_L \to\pi^0 \nu\bar{\nu})^{ }_{\rm SM} &=& (0.26 \pm 0.05) 
\times 10^{-10}~. \qquad  \label{BRKLnnt}
\eea
These results are perfectly compatible with the previous recent 
estimates reported in Ref.~\cite{rising,kettel}; 
however, it is interesting to note that the central value in the prediction 
of ${\cal B}(K^+ \to\pi^+ \nu\bar{\nu})$ has increased by $\approx 7\%$.
The main reason for this enhancement is the higher value of 
${\rm Re}\lambda_t$, resulting from a new analysis of the 
constraints imposed by $|V_{ub}|$ and $\Delta M_{B_s}$~\cite{CKMBook}.
As pointed out in Ref.~\cite{kettel}, 
the errors in Eqs.~(\ref{BRK+nnt})--(\ref{BRKLnnt})
can be reduced if ${\rm Re}\lambda_t$ and  ${\rm Im}\lambda_t$
are directly extracted from  $A_{\rm CP} (B\to J/\Psi K_S)$ and $\epsilon_K$;
however, this procedure introduces a stronger sensitivity to
the probability distribution of the (theoretical) estimate of $B_K$.
Combining the two approaches (the extraction of ${\rm Re}\lambda_t$ 
and ${\rm Im}\lambda_t$ via a global fit to the CKM matrix or
a direct extraction of ${\rm Re}\lambda_t$ and  ${\rm Im}\lambda_t$
via $A_{\rm CP} (B\to J/\Psi K_S)$ and $\epsilon_K$) 
leads to a solid upper bound of $1.0\times 10^{-10}$
on ${\cal B}(K^+ \to\pi^+ \nu\bar{\nu})^{ }_{\rm SM}$
\cite{kettel}, which represent an interesting benchmark 
for NP searches.

The high accuracy of the theoretical predictions of ${\cal B}(K^+ \to\pi^+
\nu\bar{\nu})$ and ${\cal B}(K_L \to\pi^0 \nu\bar{\nu})$ in terms of modulus
and phase of $\lambda_t=V^*_{ts} V_{td}$ clearly offers
the possibility of very interesting tests of flavour dynamics.
Within the SM, a measurement of both channels would provide 
two independent pieces of information on the unitary triangle, 
or a complete determination of $\bar\rho$ and $\bar\eta$ from $\Delta S=1$
transitions. In particular, ${\cal B}(K^+\to\pi^+ \nu\bar\nu)$
defines an ellipse in the $\bar\rho$--$\bar\eta$ plane
and ${\cal B}(K^0_{\rm L}\to\pi^0 \nu\bar\nu)$ an horizontal
line (the height of the unitarity triangle). Note, in addition, 
that the determination of $\sin 2\beta$ which can be 
obtained by combining ${\cal B}(K^0_{\rm L}\to\pi^0 \nu\bar\nu)$
and ${\cal B}(K^+\to\pi^+ \nu\bar\nu)$ is extremely clean, being 
independent from uncertainties due to $m_t$ and $V_{cb}$ (contrary 
to the separate determinations of $\bar\rho$ and $\bar\eta$)~\cite{BB3}.

\begin{figure}[t]
\vskip -0.2 cm
$$
\hskip -1.0 cm
\includegraphics[width=10.0cm,height=6.5cm]{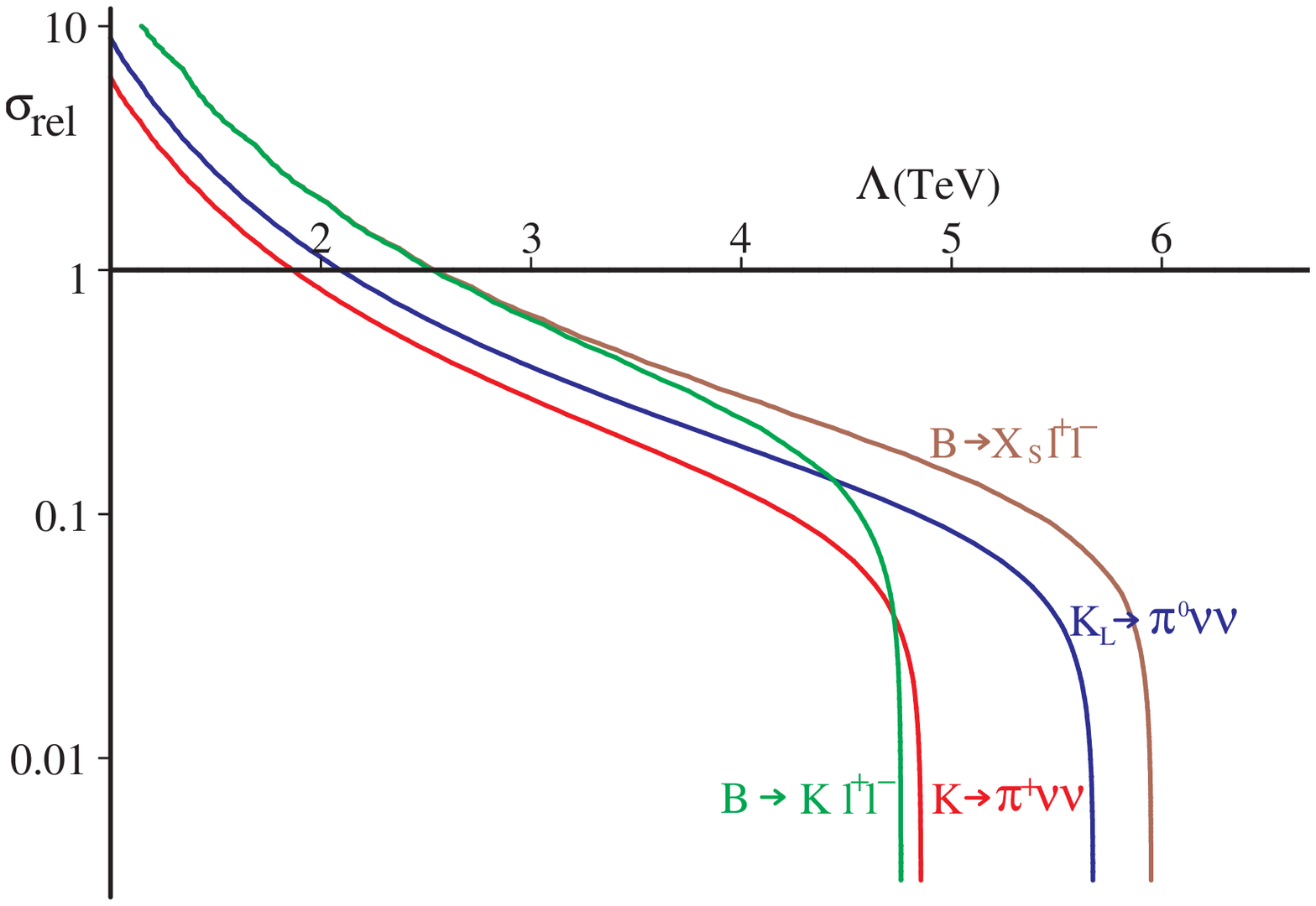} 
\hskip -1.2 cm
$$
$$
\hskip -1.0 cm
\includegraphics[width=10.0cm,height=6.5cm]{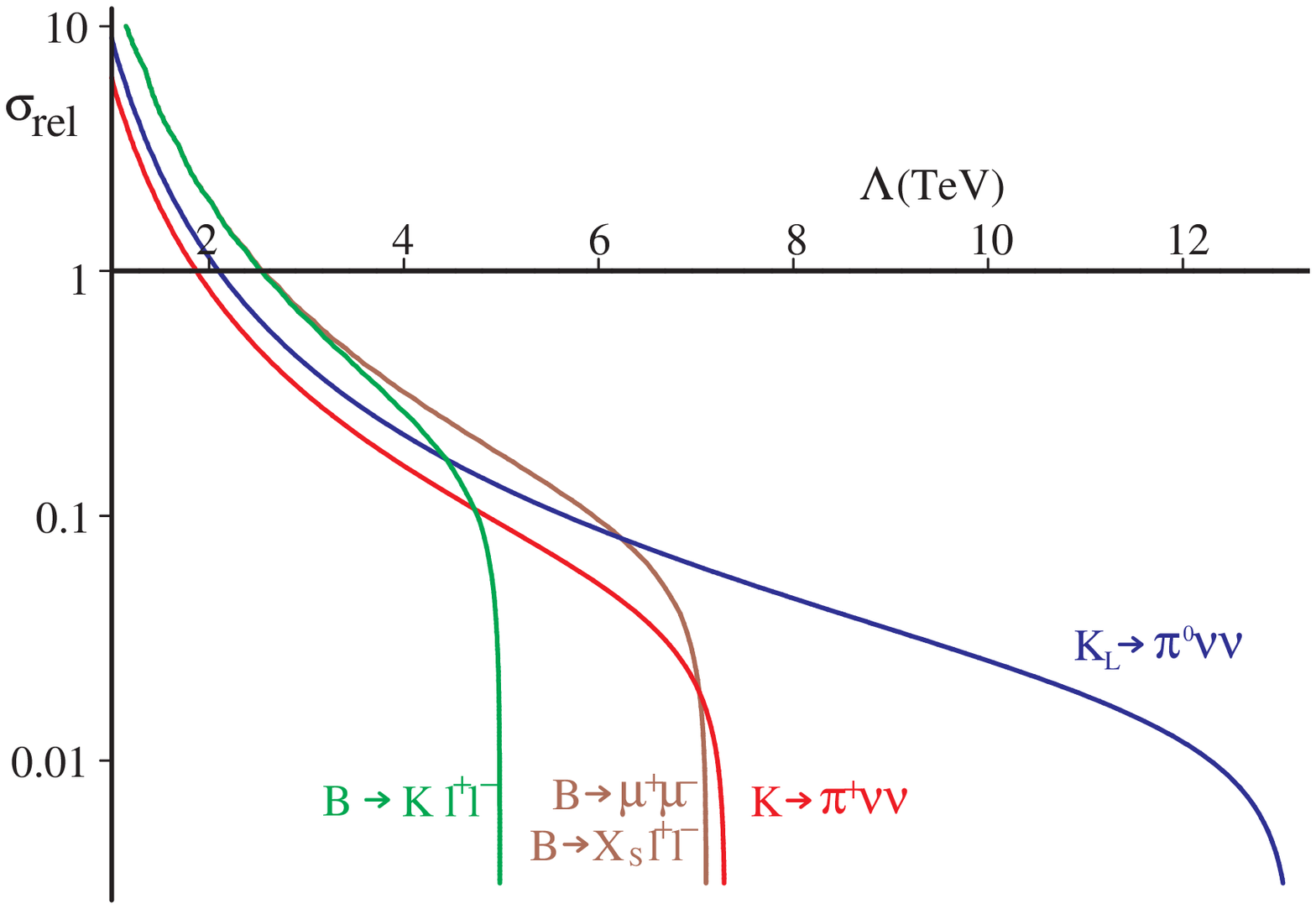}
\hskip -1.2 cm
$$
\vskip -0.8 cm
\caption[X]{\label{fig:conf} Comparison of the effectiveness 
of different rare modes  in setting future bounds 
on the scale of the representative operator 
$(\bar Q_L  \lambda_{\rm FC}\gamma_{\mu}   Q_L)(\bar L_L \gamma_\mu L_L)$
within MFV models \cite{MFV}. The vertical axis indicates the relative precision 
of an hypothetic measurement of the rate, with central value equal to
the SM expectation. The curves in the two panels are obtained
assuming an uncertainty of 10\% (left) or 1\% (right) on the
corresponding overall CKM factor. }
\end{figure}

As already mentioned, the short distance nature of the 
$s \to d \nu \bar{\nu}$ transition implies a strong 
sensitivity of $K\to \pi\nu\bar\nu$ decays to possible SM 
extensions \cite{GN}. 
Observable deviations from the SM predictions 
are expected in many specific frameworks, including low-energy 
supersymmetry \cite{Zpenguins,BCIRS}, models with extra chiral \cite{Huang} 
or vector-like quarks \cite{Hung}, and models with
large extra dimensions \cite{Buras_extra}, just to mention 
the specific frameworks which have received most of the attention 
in the last few years. Present experimental data 
do not allow yet to fully explore the high-discovery potential 
of these modes. Nonetheless, it is worth to stress that 
the evidence of the $K^+\to\pi^+ \nu\bar\nu$ transition obtained by BNL-E787 
already provides highly non-trivial constraints on the realistic scenarios 
with large new sources of flavour mixing (see e.g. Ref.~\cite{Zpenguins,GN,BCIRS}).
As illustrated in  Fig.~\ref{fig:Kpnn}, even within the pessimistic
framework of  MFV, a precise measurement of the $K_L \to \pi^0 \nu\bar\nu$ rate 
would provide --in a long term perspective--  one of the most significant constraint 
on possible new degrees of freedom.

\section{Experimental status and present impact
on CKM fits}

\begin{figure}[t] 
$$
\hskip -1.8 cm
\includegraphics[width=8.5cm,height=6.8cm]{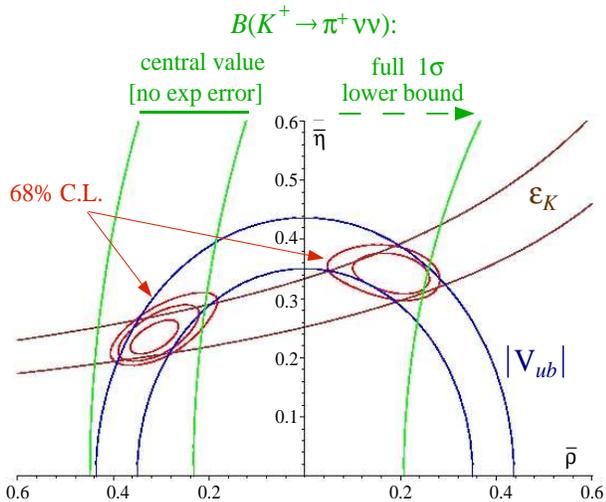}
\hskip -1.8 cm
$$
\vskip 0.1 cm
\caption{Present constraints in the $\bar\rho$--$\bar\eta$ plane 
including  $B(K^+ \to \pi^+ \nu\bar\nu)$ and 
excluding observables sensitive to $B_d$-$\bar B_d$ mixing \cite{rising}. }
\label{fig:UT_rare}
\end{figure}

The search for processes with missing energy and branching ratios 
below $10^{-10}$ is definitely a very difficult challenge, but 
has been proved not to be impossible: two $K^+ \to \pi^+ \nu\bar\nu$ 
candidate events have been observed by the BNL-E787 experiment 
\cite{E787}. The branching ratio inferred from this result,
\be
{\cal B} (K^+\to\pi^+ \nu\bar{\nu}) = \left( 1.57^{~+~1.75}_{~-~0.82} \right) \times 10^{-10}~,
\label{eq:E787}
\ee
has a central value substantially higher than the SM prediction in 
(\ref{BRK+nnt}), but is compatible with the latter once the 
large errors are take into account. In a few years this result should
be substantially improved by the BNL-E949 experiment, whose goal is to  
collect about 10 events (at the SM rate).
In the longer term, a high-precision result on this mode will arise 
from the CKM experiment at Fermilab, which aims at a  measurement 
of ${\cal B}(K^+ \to\pi^+ \nu\bar{\nu})$ at the $10\%$ level \cite{Kettel_rev}.

Unfortunately, the progress concerning the neutral mode is much slower.
No dedicated experiment has started yet (contrary to the $K^+$ case) 
and the best direct limit is more than four orders of magnitude above the 
SM expectation \cite{KTeV_nn}. An indirect model-independent upper bound  on
$\Gamma(K_L\to\pi^0\nu\bar{\nu})$ can be obtained by the isospin relation \cite{GN} 
\be
\Gamma(K^+\to\pi^+\nu\bar{\nu})~=  \Gamma(K_L\to\pi^0\nu\bar{\nu}) +
\Gamma(K_S\to\pi^0\nu\bar{\nu}) 
\label{Tri}
\ee
which is valid for any $s\to d \nu\bar\nu$ local operator of dimension 
$\leq 8$ (up to small isospin-breaking corrections).
Using the BNL-E787 result (\ref{eq:E787}), this implies 
${\cal B}(K_L\to\pi^0\nu\bar{\nu}) <  1.7 \times 10^{-9}~(90\%~{\rm CL})$.
Any experimental information below this figure can be translated into 
a non-trivial constraint on possible new-physics contributions to 
the $s\to d\nu\bar{\nu}$ amplitude. In a few years this goals should 
be reached by E931a at KEK: the first $K_L\to\pi^0\nu\bar{\nu}$ dedicated 
experiment. This experiment should eventually be upgraded in order 
to reach a SES of $10^{-13}$ and collect up to $10^3$ 
 $K_L\to\pi^0\nu\bar{\nu}$ events at the future JPARC facility \cite{JPARC}.
So far, the only approved experiment that could reach the SM 
sensitivity on $K_L\to\pi^0\nu\bar{\nu}$ is KOPIO at BNL, whose goal 
is the observation of about 50 signal events (at the SM rate) 
with signal/background~$\approx 2$ \cite{Kettel_rev}.

Although the experimental result in (\ref{eq:E787}) is not very precise 
yet, it is already quite instructive trying to use it to constrain some 
of the general NP frameworks discussed in the introduction. 
As an example, in Fig.~\ref{fig:UT_rare} we show the result of 
CKM unitarity-triangle fit allowing arbitrary NP contributions 
to $B_d$--$\bar B_d$ mixing \cite{rising}. Within this general 
frameworks, which includes one extra complex parameter with 
respect to the SM case, the standard CKM constrains 
from $A_{\rm CP} (B\to J/\Psi K_S)$, $\Delta M_{B_d}$
and $\Delta M_{B_d}/\Delta M_{B_s}$ cannot be used. In absence of the 
${\cal B}(K^+\to \pi^+\nu\bar\nu)$ information, we would find two preferred
$\bar\rho$--$\bar\eta$ regions, corresponding to the overlap 
of $\epsilon_K$ and  $|V_{ub}|$ constraints \cite{Lubicz}. 
This degeneracy persist even if we include the preliminary $A_{\rm CP}(B \to \pi^+\pi^-)$
data from $B$ factories \cite{Quim}. 
On the other hand, the ${\cal B}(K^+\to \pi^+\nu\bar\nu)$ result 
in (\ref{eq:E787}) breaks this degeneracy with a slight preference
toward the non-standard solution in the left quadrant. 
As can be seen in Fig.~\ref{fig:UT_rare},
this indication is not statistically significant yet, 
but it provides a good illustration of the main points of this discussion: 
there is still a lot to learn about
FCNC transitions and the measurements of $K \to \pi \nu\bar\nu$ rates 
provide a unique opportunity in this respect. 

\medskip

\noindent
I am grateful to Andrzej Buras for useful discussions and 
comments on the manuscript. This work is partially supported 
by IHP-RTN, EC contract No.\ HPRN-CT-2002-00311 (EURIDICE).

\end{document}